\documentclass[pra,showpacs,floatfix]{revtex4}
\usepackage{amsmath}
\usepackage{amssymb}
\usepackage{graphicx}

\begin{document}

\title{High- and low-frequency phonon modes in dipolar quantum gases trapped in
deep lattices}
\author{ Aleksandra Maluckov$^{1}$, Goran Gligori\'{c}$^{1}$, Ljup\v co Had%
\v{z}ievski$^{1}$, Boris A. Malomed$^{2}$, and Tilman Pfau$^{3}$}
\affiliation{$^1$ Vin\v ca Institute of Nuclear Sciences, University of Belgrade, P. O.
B. 522,11001 Belgrade, Serbia\\
$^{2}$ Department of Physical Electronics, School of Electrical Engineering,
Faculty of Engineering, Tel Aviv University, Tel Aviv 69978, Israel\\
$^{3}$Physikalisches Institut, Universit\"{a}t Stuttgart, Pfaffenwaldring
57, 70569 Stuttgart, Germany}

\begin{abstract}
We study normal modes propagating on top of the stable uniform background in
arrays of dipolar Bose-Einstein condensate (BEC) droplets trapped in a deep
optical lattice. Both the on-site mean-field dynamics of the droplets and
their displacement due to the repulsive dipole-dipole interactions (DDIs)
are taken into account. Dispersion relations for two modes, \textit{viz}.,
high- and low- frequency counterparts of optical and acoustic phonon modes
in condensed matter, are derived analytically and verified by direct
simulations, for both cases of the repulsive and attractive contact
interactions. The (counterpart of the) optical-phonon branch does not exist
without the DDIs. These results are relevant in the connection to emerging
experimental techniques enabling real-time imaging of the condensate
dynamics and direct experimental measurement of phonon dispersion relations
in BECs.
\end{abstract}

\maketitle

\section{Introduction}

The propagation of collective excitations in periodically structured
media has been recognized long ago as a fundamentally important
topic of condensed-matter physics \cite{Peierls}. More recently, it
has been understood that many condensed media, where the intrinsic
dynamics is very complex, may be effectively ``simulated" by
rarefied quantum gases, for which it is much easier to predict and
observe fundamental dynamical effects \cite{simulators,quant-FK}. In
the latter context, the role of the periodic structure is commonly
played by optical-lattice (OL) potentials \cite{OL}. In particular,
the use of atomic Bose-Einstein condensates (BECs) trapped in OLs
makes it possible to simulate the
propagation of acoustic waves in crystals theoretically \cite{Smerzi}-\cite%
{two-modes} and in a direct experiment \cite{exper}.

A deep OL potential splits the condensate into an array of droplets trapped
in local potential wells, which are weakly coupled with nearest neighbors by
tunneling of atoms across potential barriers. In this case, it is natural to
approximate the global BEC wave function by a superposition of wavelets of
locally trapped atoms, which resemble well-known Wannier modes that can be
used an alternative basis instead of Bloch waves \cite{Wannier}. On the
basis of this expansion, the reduction of the underlying full
Gross-Pitaevskii equation (GPE) to its discrete version (DGPE) has been
rigorously derived in several works \cite{discretization} and reviewed in
Ref. \cite{nonlinearity}. The derivation of the DGPE was further extended to
the model with a relatively strong contact nonlinearity, when the onsite
nonlinearity in the resulting discrete equation is nonpolynomial \cite{NPSE}.

A significant extension of the variety of dynamical effects in atomic BECs
is provided by long-range dipole-dipole interactions (DDIs). Basic and more
sophisticated properties of dipole condensates were predicted theoretically
\cite{theory}, which was followed by the creation of such condensates in
vapors of $^{52}$Cr \cite{experiment}. The flexibility of this setting is
buttressed by the possibility to vary the relative strength of the DDI and
contact interactions by means of the Feshbach resonance affecting the
scattering length for colliding chromium atoms \cite{Feshbach}. Results
obtained in this area have been reviewed in Ref. \cite{ddbec}.

Further progress in the experimental work has recently resulted in the
creation of dipolar condensates with the strongest possible atomic DDIs in $%
^{164}$Dy \cite{Dy}. Also promising is the work with erbium \cite{mcclelland}%
, and with gases composed of molecules carrying electric dipole moments \cite%
{hetero}.

OLs provide a powerful set of tools for the studies of dipolar BECs. In
particular, the stabilization and destabilization of the condensate, trapped
in deep OLs, by DDI was demonstrated theoretically and experimentally \cite%
{Cr-OL}. It was straightforward to generalize the DGPE for this case, which
leads to the discrete equation combining the local onsite nonlinearity with
the nonlocal nonlinear interaction between the sites, in the one- \cite%
{wannier,wannier2} and two-dimensional (2D) \cite{2Dwannier} settings alike.
Further, we have recently demonstrated that, in case the uniform (alias
continuous-wave, CW) state of the dipolar condensate, trapped in a deep
periodic potential (i.e., the discretized state with equal amplitudes at all
sites of the lattice), is modulationally unstable, two- and three-period
density patterns emerge as energy minimizers, the stability area being very
large for the three-period pattern \cite{prlnash}. Modulated-density
solutions are also produced by the DGPE for condensates fragmented by the
deep OL potential in the absence of long-range interactions, but in that
case they are completely unstable \cite{Nicolin}. 2D supersolid structures
supported by the OL in the dipolar BEC were predicted too \cite{BuBu}, and
anisotropic DDIs may give rise to striped 2D\ patterns even in the absence
of the lattice \cite{stripes}.

Beyond the aforementioned detailed studies of static patterns, a physical
problem of obvious interest is the study of collective excitations
propagating on top of stable states supported by the long-range
interactions. In this context, a well-known result is the prediction of the
roton branch of excitations in the dipolar condensates \cite{roton}. The
subject of the present work is the analysis of excitation modes propagating
through the dipolar BEC fragmented by the deep OL and, accordingly, modeled
by the DGPE including the long-range DDIs. The rapid progress in
experimental studies of condensates (in particular, the development of
techniques for in-situ imaging of the condensate dynamics \cite%
{insitu,shammas}) suggests that the direct observation of such modes will
become possible, hence the analysis is a subject of direct relevance. A
related promising experimental technique, based on the method of the Bragg
spectroscopy \cite{Bragg}, was very recently elaborated for directly
measuring phonon dispersion relations in trapped BECs \cite{shammas}. In
this work, we focus on the most fundamental case of the normal modes excited
on top of the stable CW state. Still more challenging situations with the
underlying periodically modulated density profiles will be considered
elsewhere.

The most essential result of the analysis is that the long-range DDI between
condensate droplets gives rise to a high-frequency branch of collective
excitations, which emulates the optical-phonon mode well known in
condensed-matter physics. This excitation propagates as a displacement wave
in the array of trapped BEC droplets, which resembles the dynamics of the
Frenkel-Kontorova model \cite{FK}. Another fundamental branch of the
propagating waves represents low-frequency density oscillations of the
droplets, which is a counterpart of acoustic phonons in condensed matter. It
is a generalization of the branch which was recently found in the limit of
immobile droplets \cite{prlnash}.

Excitation modes similar to those reported here may be expected in other
physical settings which feature long-range interactions between elements of
a fragmented wave field. In particular, laser illumination of atoms may
induce an effective gravity-like long-range attraction in BEC, which, in the
combination with the OL, may give rise to a supersolid \cite{Gershon}. An
example in optics is offered by the light field trapped in an arrayed
waveguide embedded into a medium with the nonlocal thermal nonlinearity \cite%
{thermal}, which gives rise to a number of patterns in one and two
dimensions, as demonstrated theoretically \cite{Krolik} and experimentally
\cite{Segev}. Other possible realizations\ of this generic setting in optics
may use the nonlocal nonlinearity of liquid crystals \cite{LC}, and an
effective nonlocality in semiconductor waveguides induced by the transport
of charge carriers \cite{Lederer}.

\section{The model}

\begin{figure}[h]
\centering
\includegraphics[width=10cm]{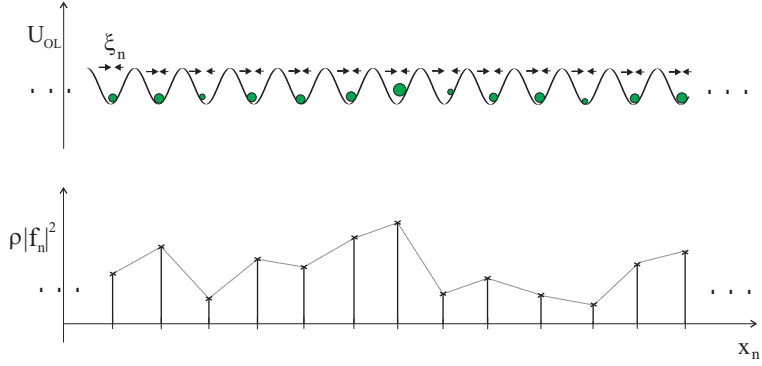}
\caption{(Color online) In the top panel, circles represent the array of
droplets trapped in the OL potential. The bottom panel displays a generic
distribution of droplet masses, which are proportional to areas of the
circles.}
\label{fig0}
\end{figure}

The on-site dynamics of the dipolar-BEC droplets trapped in local potential
wells of the deep OL potential is governed by the DGPE, which includes the
onsite cubic nonlinearity and long-range DDI. Following the notation adopted
in Ref. \cite{prlnash}, the discrete equation is written as
\begin{equation}
i\frac{df_{n}}{dt}=-C_{n}(f_{n+1}-f_{n})-C_{n-1}(f_{n-1}-f_{n})+\sigma
\left\vert f_{n}\right\vert ^{2}f_{n}-\Gamma \sum_{m\neq n}\frac{\left\vert
f_{m}\right\vert ^{2}}{|x_{m}-x_{n}|^{3}}f_{n},  \label{eq1}
\end{equation}%
where $f_{n}$ is the discrete wave function, $C_{n}$ are inter-site coupling
coefficients proportional to the overlapping integrals between the Wannier
modes exponentially localized at adjacent sites \cite{discretization}-\cite%
{wannier}. If the coordinate of the $n$-th droplet is taken as $x_{n}=n+\xi
_{n}$, where $\xi _{n}$ accounts for its displacements due to the DDIs with
other droplets (see Fig. \ref{fig0}), then the corresponding dependence of
coefficients $C_{n}$ in Eq. (\ref{eq1}) on $\xi _{n}$ is
\begin{equation}
C_{n}=C^{(0)}\exp \left( -\frac{1+\xi _{n+1}-\xi _{n}}{l}\right) \equiv
c\exp \left( -\frac{\xi _{n+1}-\xi _{n}}{l}\right) ,  \label{C}
\end{equation}%
$c\equiv C^{(0)}\exp \left( -1/l\right) $, and $C^{(0)}$ is a coefficient of
tunneling of atoms between adjacent sites separated by distance $l$.

Proceeding to nonlinear terms in Eq. (\ref{eq1}), $\sigma =-1$ and $+1$
correspond to the attractive or repulsive onsite contact interaction
(abbreviated below as AC or RC, respectively), and $\Gamma $ is the relative
strength of the DDI, with respect to the onsite nonlinearity \cite{wannier}.
In the present notation, $\Gamma >0$ and $\Gamma <0$ correspond to the
attractive and repulsive DDIs, respectively. According to its definition, $%
\Gamma $ may be adjusted by means of the Feshbach resonance which affects
the strength of the local nonlinearity, and, additionally, by fixing the
orientation of the magnetic moments with respect to the system's axis.

The energies of the DDI, and of the trapping of the condensate in the OL
potential are
\begin{equation}
U_{\mathrm{DD}}=-\Gamma \sum_{m\neq n}\frac{\left\vert f_{m}\right\vert
^{2}\left\vert f_{n}\right\vert ^{2}}{\left\vert x_{n}-x_{m}\right\vert ^{3}}%
,~U_{\mathrm{OL}}=-U_{0}\sum_{n}\cos \left( 2\pi x_{n}\right) ,  \label{UDD}
\end{equation}%
where $U_{0}$ is the depth of the periodic potential, hence the motion of
each droplet is governed by the Newton's equation,
\begin{equation}
\rho \frac{d}{dt}\left( \left\vert f_{n}\right\vert ^{2}\frac{d\xi _{n}}{dt}%
\right) =-\frac{\partial }{\partial \xi _{n}}\left( U_{\mathrm{DD}}+U_{%
\mathrm{OL}}\right) ,  \label{Newton}
\end{equation}%
where $\rho $ is the density of the condensate, so that the mass and
momentum of the $n$-th droplet are $\rho \left\vert f_{n}\right\vert ^{2}$
and $\rho \left\vert f_{n}\right\vert ^{2}d\xi _{n}/dt$, as shown in Fig. %
\ref{fig0}.

Thus, the dynamical system following from Eqs. (\ref{eq1}), (\ref{Newton})
and (\ref{UDD}) takes the following form:
\begin{gather}
i\frac{df_{n}}{dt}=-c\left[ e^{-\frac{\xi _{n+1}-\xi _{n}}{l}%
}(f_{n+1}-f_{n})+e^{-\frac{\xi _{n}-\xi _{n-1}}{l}}(f_{n-1}-f_{n})\right]
\notag \\
+\sigma \left\vert f_{n}\right\vert ^{2}f_{n}-\Gamma \sum_{m\neq n}\frac{%
\left\vert f_{m}\right\vert ^{2}}{|n-m+\xi _{n}-\xi _{m}|^{3}}f_{n}~,
\label{eq10} \\
\rho \frac{d}{dt}\left( \left\vert f_{n}\right\vert ^{2}\frac{d\xi _{n}}{dt}%
\right) =-2\pi U_{0}\sin (2\pi \xi _{n})  \notag \\
+3\Gamma \left\vert f_{n}\right\vert ^{2}\left( \sum_{m>n}-\sum_{m<n}\right)
\left( \frac{\left\vert f_{m}\right\vert ^{2}}{|n-m+\xi _{n}-\xi _{m}|^{4}}%
\right) ~.  \label{eq11}
\end{gather}%
This system was simulated by means of the Runge-Kutta method of the sixth
order.

\section{Normal excitation modes}

To study collective-excitation modes in the present context, we set $%
f_{n}=(\phi _{0}+g_{n})e^{-i\mu t}$, where $\phi _{0}=\sqrt{\mu /\left[
\sigma -2\Gamma \zeta (3)\right] }$ is the amplitude of the CW state, as
found in Ref. \cite{prlnash} [$\zeta (3)\approx \allowbreak 1.2021$ is the
value of the Riemann's zeta-function], and linearize Eqs. (\ref{eq10}) and (%
\ref{eq11}) with respect to small perturbations, $g_{n}$ and $\xi _{n}$:
\begin{gather}
i\frac{dg_{n}}{dt}=\left[ -\mu +2c+2\left( \sigma -2\Gamma \zeta (3)\right)
\phi _{0}^{2}\right] g_{n}-c\left( g_{n+1}+g_{n-1}\right)  \notag \\
-\Gamma \phi _{0}^{2}\sum_{q>0}q^{-3}\Pi _{q}+\sigma \phi
_{0}^{2}g_{n}^{\ast }+3\phi _{0}^{3}\Gamma \sum_{q>0}q^{-4}\left( \xi
_{n+q}-\xi _{n-q}\right) ~,  \notag \\
\rho \frac{d^{2}\xi _{n}}{dt^{2}}=\left[ -\left( \frac{2\pi }{\phi _{0}}%
\right) ^{2}U_{0}-24\Gamma \phi _{0}^{2}\sum_{q>0}q^{-5}\right] \xi _{n}
\notag \\
+24\Gamma \phi _{0}^{2}\sum_{q>0}q^{-5}\left( g_{n-q}^{\ast }-g_{n+q}^{\ast
}+g_{n-q}^{\ast }-g_{n+q}^{\ast }\right) ~,  \label{eqn}
\end{gather}%
where $q$ is an integer index of the summation over all sites of the
lattice, the asterisk stands for the complex conjugate, and $\Pi _{q}\equiv
g_{n-q}+g_{n+q}+g_{n-q}^{\ast }+g_{n+q}^{\ast }$. Difference-differential
equations (\ref{eqn}) are Fourier-transformed into the system of algebraic
equations,
\begin{equation}
\left[
\begin{array}{ccc}
A+\omega & -B & -iF \\
-B & A-\omega & -iF \\
i\left( \rho \phi _{0}\right) ^{-1}F & i\left( \rho \phi _{0}\right) ^{-1}F
& D-\omega ^{2}%
\end{array}%
\right] \left[
\begin{array}{c}
G_{k}(\omega ) \\
G_{-k}^{\ast }(-\omega ) \\
\Xi _{k}(\omega )%
\end{array}%
\right] =0  \label{eq00}
\end{equation}%
where $G_{k}(\omega )$ and $\Xi _{k}(\omega )$ are the Fourier transforms of
$g_{n}(t)$ and $\xi _{n}(t)$, and
\begin{eqnarray}
A &\equiv &-4c\sin ^{2}\left( \frac{k}{2}\right) -B,~B\equiv \phi _{0}^{2}%
\left[ \sigma -2\Gamma \sum_{q>0}\frac{\cos (kq)}{q^{3}}\right] ,  \notag \\
F &\equiv &6\Gamma \phi _{0}^{2}\sum_{q>0}\frac{\sin (kq)}{q^{4}},  \notag \\
D &\equiv &\frac{1}{\rho }\left\{ \left( \frac{2\pi }{\phi _{0}}\right)
^{2}U_{0}+24\Gamma \phi _{0}^{2}\left[ \zeta (5)-\sum_{q>0}q^{-5}\cos (kq)%
\right] \right\} ,  \label{kkk}
\end{eqnarray}%
with $\zeta (5)\approx \allowbreak 1.0369$.

Dispersion relations for propagating waves, $\omega =\omega (k)$, are
derived from Eq. (\ref{eq00}) by equating the determinant of the system's
matrix to zero, which yields a biquadratic equation,
\begin{equation}
\omega ^{4}-(A^{2}-B^{2}+D)\omega ^{2}+D(A^{2}-B^{2})-\frac{2F^{2}}{\rho
\phi _{0}}(A+B)=0.  \label{dispersion}
\end{equation}%
In the parameter region of interest, which corresponds to the stable CW
background, Eq. (\ref{dispersion}) yields four solutions for the frequency, $%
\omega (k)=\left\{ \omega _{1}(k),\omega _{2}(k)=-\omega _{1}(k),\omega
_{3}(k),\omega _{4}(k)=-\omega _{3}(k)\right\} $. In the following we
display the positive-frequency branches, the negative ones being their
specular images. Further straightforward analysis of the solutions produced
by Eq. (\ref{eq00}) demonstrates that the low- and high-frequency branches
represent waves of onsite density oscillations and droplet displacements,
respectively. The latter one may also be called a \textit{dipolar-phonon}
mode.

The BEC system which does not admit the droplet displacements (in other
words, the corresponding trapping frequency is assumed to be infinite) is
modeled by the first equation in (\ref{eq11}) with $\xi _{n}\equiv 0$ \cite%
{prlnash}. The corresponding dispersion relation amounts to the single
positive real root for $\omega ^{2}$ in the parameter region where the CW
background is stable. It represents the density-oscillation mode in the case
of frozen displacements.

The modes in the present system may be compared to Langmuir and
ion-acoustic waves in plasmas \cite{plasma}. In this context, the
displacement mode, although it originates as the dipolar-phonon
wave, is a counterpart of the high-frequency Langmuir wave of
electron oscillations in the plasma, while the density-oscillation
mode is an analog of the ion sound. The reason for this ``switch of
the roles" is that the displacement wave is
actually a stream of \textit{optical phonons}, in terms of solid state \cite%
{Peierls}, while the ion sound in the plasma is carried by \textit{acoustic
phonons}. Further, in the long-wave limit the dispersion relation for the
low-frequency density mode takes precisely the form of the $\omega (k)$
relation for acoustic phonons, see Eq. (\ref{nu}) below.

\section{Results and discussion}

\subsection{Analytical considerations}

The analysis of stability conditions for the CW background in the present
system demonstrates that they are identical to those in the previously
considered model \cite{prlnash} with the frozen displacement degree of
freedom. For the RC interactions [$\sigma =+1$ in Eq. (\ref{eq1})], the
CW-stability region extends to arbitrarily weak repulsive DDIs, $\Gamma \leq
0$ (see Fig. 1 in Ref. \cite{prlnash}), while in the case of the AC
interactions ($\sigma =-1$) the CW has a stability region if the repulsive
DDI is strong enough,
\begin{equation}
\left\vert \Gamma \right\vert >\left\vert \Gamma _{\mathrm{thr}}\right\vert ,
\label{thr}
\end{equation}%
with the threshold value that can be found from dispersion relation (\ref%
{dispersion}). For $\mu =1$ (the chemical potential which is adopted in
figures displayed here), $\Gamma _{\mathrm{thr}}\approx \allowbreak -1.27$.
To relate these results to the experimental situation, it is relevant to
mention that as shown in Ref. [13], a typical value $\left\vert \Gamma
\right\vert =0.5$, relevant to our analysis (see Fig. \ref{fig1} below),
corresponds to parameter values which are realistic in experiments with $%
^{52}$Cr: the scattering length $\simeq 2$ nm and the transverse-confinement
radius $\simeq $ $5$ mm. Actually, positive and negative values of $\Gamma $
may be varied within broad limits by means of the Feshbach resonance, as
shown experimentally \cite{experiment}.

As mentioned above, the analytical solution of dispersion equation (\ref%
{dispersion}) demonstrates that the stable CW background supports two modes
of the propagating excitations, \textit{viz}., the low-frequency
acoustic-phonon and high-frequency optical-phonon modes. The two dispersion
curves are displayed in Fig. \ref{fig1}, for both signs of the contact
interaction, RC and AC. A noteworthy peculiarity of the latter (AC) case is
that the dispersion relation has a minimum at the center of the Brillouin
zone, provided that the repulsive DDI is relatively weak. We stress that
only the situations with the stable CW background are shown, therefore
values of $\Gamma $ which do not satisfy stability condition (\ref{thr}) are
not included in panel \ref{fig1}(b).

\begin{figure}[tbp]
\centering
\includegraphics[width=12cm]{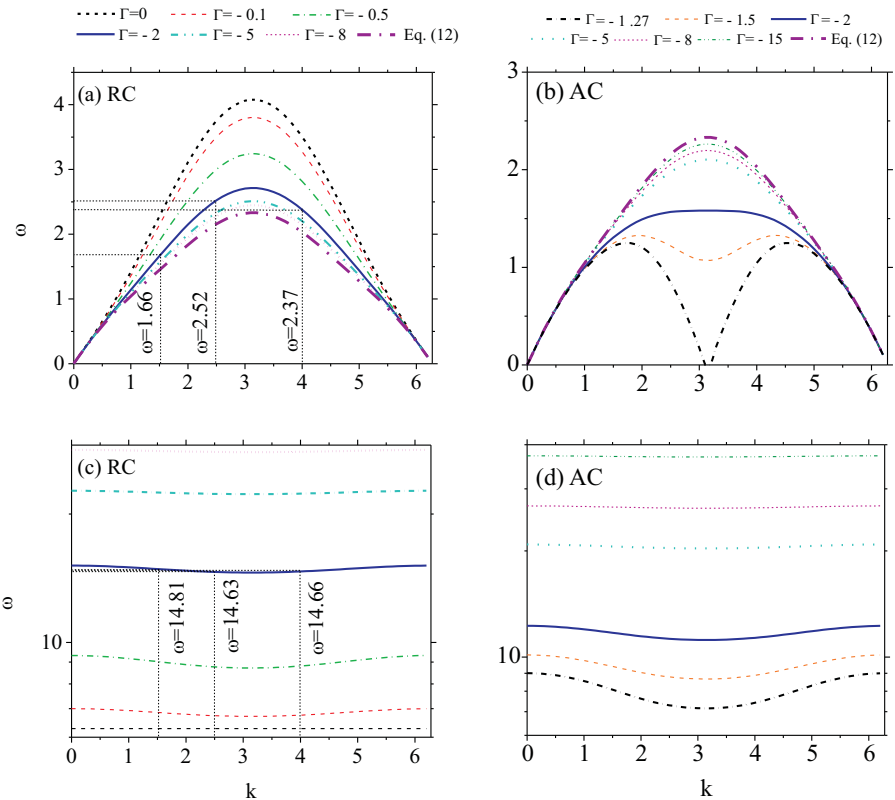}
\caption{(Color online) The top row: dispersion curves for the low-frequency
acoustic phonon mode, with repulsive (a) and attractive (b) signs of the
contact interaction (RC and AC, respectively). The bottom row: dispersion
curves for the high-frequency optical-phonon mode, with the RC (c) and AC
(d) signs of the local nonlinearity. The curves are drawn for $\protect%
\mu =1$, in the first Brillouin zone, $0\leq k<2\protect\pi $. The
corresponding relative values of the DDI strength, $\Gamma $, are indicated
in the plots. Vertical lines in panels (a) and (c) indicate frequencies
predicted by the two dispersion curves at three wavenumbers, $k=1.57,$ $2.5$%
, and $4$, for which numerically computed temporal spectra of lattice
perturbations, excited as per Eq. (\protect\ref{in}), are displayed below in
Fig. \protect\ref{fig3}. }
\label{fig1}
\end{figure}

The decrease of the largest frequency of the acoustic-phonon mode
with the increase of $|\Gamma |$, which is observed in Fig.
\ref{fig1}(a), implies that the system becomes ``less stiff", for
the propagation of the density waves, with the strengthening of the
repulsive DDI. A reason for this counter-intuitive conclusion is
that the model does not directly take into account the DDI inside
each droplet. However, this may be included via a renormalization of
the onsite contact interaction\textit{.} On the other hand, Fig.
\ref{fig1}(b) demonstrates that the stiffness increases with
$|\Gamma |$ in the case of the AC interactions ($\sigma =-1$). In
either cases of the RC or AC interactions, with the increase of
$|\Gamma |$ the dispersion curves for the low-frequency (LF)
acoustic-phonon mode
asymptotically approach a limit form which can be derived from Eqs. (\ref%
{kkk}) and (\ref{dispersion}):
\begin{equation}
\omega _{\mathrm{LF}}(\Gamma \rightarrow -\infty )=\pm 4c\sin \left( \frac{k%
}{2}\right) \sqrt{\sin ^{2}\left( \frac{k}{2}\right) +\frac{\mu }{2c\zeta (3)%
}\sum_{q>0}q^{-3}\cos (kq)}.  \label{asympt}
\end{equation}%
In Figs. \ref{fig1}(a) and (b), the dispersion curves approach the
asymptotic one from above and below in the cases of the RC and AC
interactions, respectively.

Naturally, these results imply that the acoustic-phonon mode can propagate
in the absence of the DDI ($\Gamma =0$) in the condensate with the RC
interactions only. The addition of the DDIs modifies the corresponding
dispersion curve without altering it qualitatively. Note also that the
acoustic-phonon mode curve takes the form of $\omega \approx vk$ for small $%
k $, with the effective sound velocity that can be found from Eqs. (\ref%
{dispersion}) and (\ref{kkk}):%
\begin{equation}
v=\phi _{0}\sqrt{2c\left( \sigma -2\Gamma \zeta (3)\right) }.  \label{nu}
\end{equation}

On the other hand, the optical-phonon mode cannot propagate in the absence
of the repulsive DDI, i.e., at $\Gamma =0$, when it degenerates into
uncoupled oscillations of the droplets around local minima of the lattice
potential, with frequency $\omega _{\mathrm{osc}}=\left( 2\pi /\phi
_{0}\right) \sqrt{U_{0}/\rho }$ [see the horizontal black dotted line in
Fig. \ref{fig1} (c)]. Further, the expansion of the $\omega (k)$ relation
for the high-frequency (HF) optical-phonon mode in the long-wave limit (for
small $k^{2}$) exhibits a quadratic dispersion, similar to that for the
Langmuir waves in the plasma:
\begin{equation}
\omega _{\mathrm{HF}}(k)\approx \frac{2\pi }{\phi _{0}}\sqrt{\frac{U_{0}}{%
\rho }}-\frac{3\phi _{0}^{3}\zeta (3)}{\pi }\frac{\left\vert \Gamma
\right\vert }{\sqrt{U_{0}\rho }}k^{2}.
\end{equation}

\subsection{Direct simulations}

\begin{figure}[tbp]
\centering
\includegraphics[width=12cm]{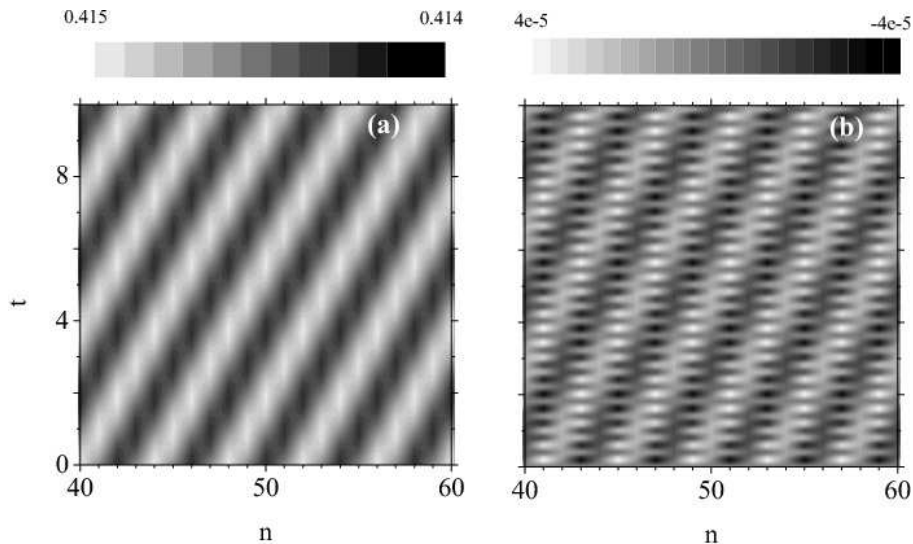}
\caption{(Color online) Contour plots of the density, $|f_{n}|$ (a), and
displacement, $\protect\xi _{n}$ (b), for simultaneously excited acoustic
and optical phonon waves, in the case of the repulsive contact and
dipole-dipole interactions, $\protect\sigma =+1$ and $\Gamma =-2$. The
wavenumber and amplitude of the initial excitation (\protect\ref{in}) are $k=%
\protect\pi /2=1.57$ and \textbf{\ }$\protect\varepsilon =0.001$. The
respective frequencies of the propagating low- and high-frequency modes are $%
\protect\omega _{\mathrm{LF}}=1.8$ (a), and $\protect\omega _{\mathrm{HF}%
}=14.9$ (b), which agree with those obtained from dispersion relations (%
\protect\ref{dispersion}) and the temporal spectra (see Fig. \protect\ref%
{fig3}). }
\label{fig2}
\end{figure}

The propagation of the waves through the stable CW background was tested by
direct simulations of Eqs. (\ref{eqn}). The system was initially excited by
a perturbation in the form of
\begin{equation}
f_{n}(t=0)=\phi _{0}+\varepsilon \exp (ikn),  \label{in}
\end{equation}%
with small amplitude $\varepsilon $ and arbitrary wavenumber $k$. The
simulations demonstrate a superposition of the acoustic-phonon and
optical-phonon propagating modes, as shown in Fig. \ref{fig2}.

\begin{figure}[tbp]
\centering
\includegraphics[width=12cm]{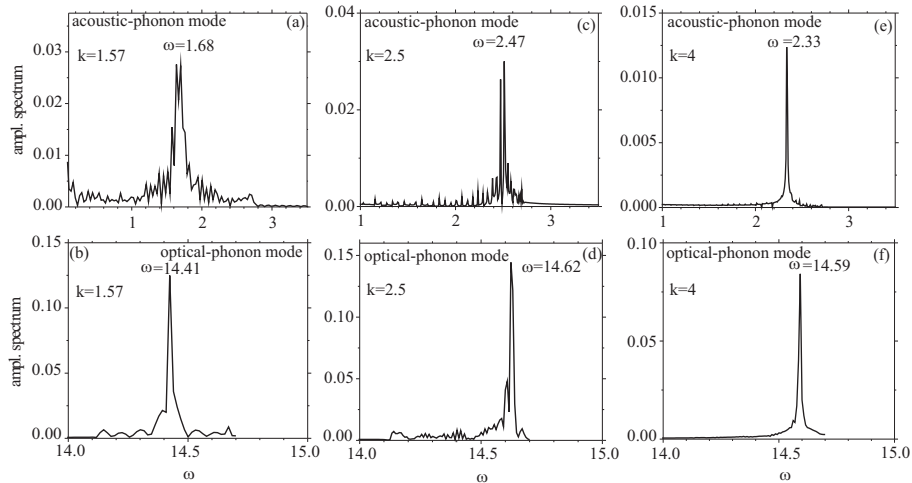}
\caption{Typical examples of temporal spectra of excitations in the system
with $\protect\sigma =+1$, $\Gamma =-2$, $\protect\mu =1$, computed at a
particular site of the lattice: (a), (c), (e) -- the acoustic-phonon modes;
(b), (d), (f) -- the optical-phonon modes. The peak values of $\protect%
\omega $ accurately fit their analytical counterparts obtained from Eq. (%
\protect\ref{dispersion}), which are marked in Fig. \protect\ref{fig1}.}
\label{fig3}
\end{figure}

Characteristic frequencies of both types of the propagating modes, as
obtained from the direct simulations, almost precisely fit those predicted
by the dispersion relations. This is shown in Fig. \ref{fig3} by means of
sample temporal spectra representing the propagating perturbations excited
as per Eq. (\ref{in}) with \textbf{\ }$\varepsilon =0.001$ and three
different wavenumbers, $k=1.57,~2.5,$ and $4$, in the case of the RC
interactions. For each $k$, the creation of the acoustic-phonon and
optical-phonon modes is observed. The central frequencies, indicated in the
plots, are very close to their counterparts predicted by the dispersion
curves, which are marked by vertical lines in Figs. \ref{fig1}(a,c).

Additional simulations (not displayed here) demonstrate that the increase of
the amplitude of the initial perturbation ($\varepsilon $) naturally makes
the spectra wider and more complex.

\section{Conclusions}

This work reports a new mode of high-frequency collective excitations
propagating on top of the stable background in the dipolar BEC trapped in
the form of the array of droplets in a deep OL potential. This branch of the
excitations, which is similar to the optical-phonon mode in condensed
matter, and also features similarity to Langmuir waves in plasmas, does not
exist in the absence of the repulsive dipole-dipole interactions. The
dispersion relations for the new mode and the additional low-frequency one
of the acoustic-phonon type, which accounts for density oscillations in the
trapped BEC, were derived analytically. The results have been confirmed by
direct simulations of the excited system.

The currently developing techniques for in situ imaging of dynamical
phenomena and recording phonon dispersion relations in trapped BEC \cite%
{insitu,shammas} open the way to experimental realization of the results
reported in this work in atomic gases of chromium, dysprosium and erbium,
and in molecular gases, such as KRb and NaLi. As concerns the theoretical
analysis, it can be extended in several directions. First, it would be
interesting to study the nonlinear dynamics of excitations with nonsmall
amplitudes. Second, a natural extension is to consider collective
excitations on top of stable periodically modulated patterns, such as the
double- and triple-periodic ones \cite{prlnash}. This extension is expected
to produce a larger number of different dynamical branches. Then, a
challenging problem is to find normal modes of collective excitations in 2D
settings. Finally, quite an interesting possibility is to introduce a
deformable Bose-Hubbard model, as a quantum version of the mean-field system
considered in the present work, which may be then implemented in chains of
trapped ultracold atoms or ions \cite{quant-FK}.

A.M., G.G., and Lj.H. acknowledge support from the Ministry of Education and
Science of Serbia (Project III45010). The work of B.A.M. and T.P. was
supported in a part by the German-Israel Foundation through grant No.
I-1024-2.7/2009.

\end{document}